# Nonlinear vibrational spectrometer for bioapplications featuring narrowband 1-μm pulses and a recycled OPA pump beam

Zsuzsanna Heiner,[1,*] Valentin Petrov,[2] and Mark Mero[2]

[1] School of Analytical Sciences Adlershof, Humboldt-Universität zu Berlin, 12489 Berlin, Germany
[2] Max Born Institute for Nonlinear Optics and Short Pulse Spectroscopy, 12489 Berlin, Germany
* zsuzsanna.heiner@hu-berlin.de

**Abstract:** Moving the detection wavelength in vibrational sum-frequency generation (VSFG) spectroscopy to the near-infrared (> 700 nm) can potentially enable the study of molecular interfaces absorbing in the visible and give access to buried bio-interfaces at minimal absorption, reduced scattering, and negligible autofluorescence. Here, we employ an ultra-narrow bandpass thin-film optical interference filter on 180-fs, 1.03-μm laser pulses to generate an upconversion beam yielding a spectral resolution of 5 cm$^{-1}$ and VSFG wavelengths between 890 and 980 nm for molecular vibrations in the fingerprint region. We demonstrate that the beam rejected by the filter can be utilized for driving a supercontinuum-seeded near-infrared optical parametric amplifier serving as the front-end of a broadband LiGaS$_2$-based mid-infrared amplifier. Benchmark data on a phospholipid monolayer at the air-water interface acquired using the resulting VSFG spectrometer show the possibility of achieving high resolution and signal-to-noise ratio at short acquisition times. The scheme can also be utilized in other types of vibrational spectroscopy that derive their spectral resolution from bandpass-filtering of femtosecond near-infrared laser pulses, such as stimulated Raman scattering (SRS) and coherent anti-Stokes Raman scattering (CARS) spectroscopy.

1. **Introduction**

Broadband vibrational sum frequency generation (VSFG) spectroscopy is a nonlinear optical technique for acquiring the vibrational spectrum of interfacial molecular layers without a background contribution from a centrosymmetric bulk phase [1, 2]. The VSFG process relies on three-wave mixing, where photons from a broadband mid-infrared (MIR, 3-30 μm) laser beam resonant with the molecular vibrations to be studied are mixed with photons from nonresonant, narrowband, so-called visible (VIS) laser pulses that serve as the frequency upconversion beam. The spectral resolution in various types of nonlinear vibrational spectroscopy, including VSFG spectroscopy, is limited by the spectral bandwidth of the narrowband pulses. Over the years, considerable effort has been dedicated to producing these pulses with sufficiently narrow bandwidth. Historically, the second harmonic of nanosecond Nd:YAG lasers was first used as the narrowband upconversion beam [1], thus the designation, visible. However, with the spread of Ti:sapphire lasers, the duration and wavelength of the narrowband beam moved to the picosecond and near-infrared (NIR) region, respectively. For this reason, we will use the term picosecond near-infrared (ps-NIR) for upconversion beams with wavelengths above 700 nm.

Picosecond VIS/NIR pulses are commonly generated using either spectral filtering or nonlinear frequency upconversion of the femtosecond laser pulses driving the VSFG spectrometer [3]. Less common schemes rely on a second, narrowband laser amplifier channel synchronized to the mid-infrared channel [4, 5]. In spectral filtering using bandpass filters, etalons, and 4f shapers, most of the energy of the input laser beam is discarded, leading to high losses that can only partially be compensated by the amplification of the filtered beam [6]. The transmitted beam is centered at the fundamental wavelength of either 800 or 1030 nm in the case of Ti:sapphire or Yb lasers, respectively. With Ti:sapphire systems (high pulse energy), such losses are tolerable, but with multi-10-kHz repetition rate Yb laser systems operating at moderate pulse energies, this is often too costly. Typically, the bandwidths near 800 nm are limited to > 10 $cm^{-1}$ with bandpass filters and standard etalons [7]. An alternative, much more efficient way is to rely on frequency doubling in long crystals [8] or so-called spectral compressors, where chirped sum-frequency generation is used to arrive also at approximately the second harmonic wavelength [9, 10]. The center wavelength of the VIS beam is ~515 nm in this case, which leads to VSFG wavelengths in the range of ~430-500 nm for vibrational wavenumbers in the range of ca. 500-4000 $cm^{-1}$ (cf. the range of 500-1500 $cm^{-1}$ constitutes the molecular fingerprint region). While spectral compression can generate bandwidths down to 0.16 $cm^{-1}$ at an attractively high conversion efficiency of 26% [11], electronic resonance effects and re-absorption of the generated short-wavelength SFG photons can occur for a wide range of biological samples, such as autofluorescent molecules.

In this work, we leverage (i) a commercially widely available high-performance thin-film optical interference filter for ultra-narrow bandpass filtering at the fundamental of an 1028-nm Yb:KGd(WO$_4$)$_2$ laser to generate a ps-NIR beam with a bandwidth of 5 $cm^{-1}$ and (ii) recycling of the beam reflected off the bandpass filter to drive a supercontinuum-seeded NIR optical parametric amplifier (OPA) front-end for seeding a broadband MIR LiGaS$_2$ (LGS)-based OPA operating in the molecular fingerprint region. By reusing the back-reflected beam rejected by the bandpass filter, which is usually discarded in similar schemes, we were able to significantly boost the overall efficiency of our VSFG spectrometer. Despite the tacit assumption that femtosecond pulses reflected off a narrow bandpass filter must necessarily be unsuitable for further use, we demonstrate that such a beam can be used to drive a supercontinuum-seeded femtosecond OPA where both the signal and idler output beams are of high spatio-temporal quality with smooth polynomial spectral phases. Thanks to the long-wave ps-NIR beam, the VSFG signal falls in the ~730-980 nm range for vibrational wavenumbers of 500-4000 $cm^{-1}$, enabling the study of sensitive biological samples.

## 2. Experimental setup

The scheme of the light source driving the VSFG spectrometer is shown in Fig. 1. Pulses with an energy of 60 µJ delivered from a commercial Yb:KGd(WO$_4$)$_2$ laser (Pharos SP, Light Conversion Ltd.) operating at 1028-nm center wavelength were used to pump the entire system at a repetition rate of 100 kHz. For the upconversion beamline, a 2.0-W portion with 20 µJ/pulse was separated with a partial reflector (cf. PR in Fig. 1) and was spectrally filtered with an ultra-narrow bandpass filter (Alluxa, Inc., cf. BP in Fig. 1), specified with a full width at half maximum (FWHM) bandwidth of < 0.7 nm at 1031.20 nm for 0° angle of incidence, to produce narrowband, ps-NIR pulses from the 180-fs incident pump pulses. The power transmitted through the filter at our center wavelength of 1028 nm was maximized by optimizing the tilt angle of the filter, leading to an energy of 0.86 µJ/pulse.

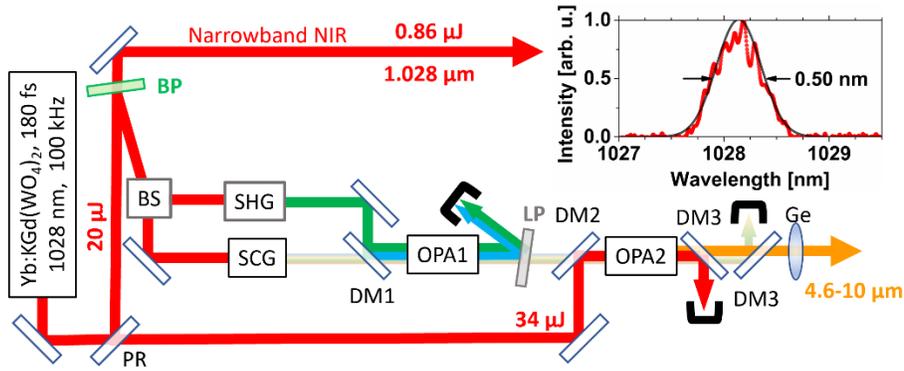

**Fig. 1.** Schematic layout of the light source driving the VSFG spectrometer. PR: partial reflector, BS: beam splitter, SCG: supercontinuum generation unit, SHG: second-harmonic generation unit, OPA1 and OPA2: optical parametric amplifiers, DM1: dichroic mirror, high reflection at 515 nm and high transmission at > 1.1 µm, DM2: dichroic mirror, high reflection at 1028 nm and high transmission at > 1.1 µm, DM3: dichroic mirror, high reflection at 1-1.25 µm and high transmission above 4.5 µm, BP: ultra-narrow bandpass filter, LP: longpass thin-film interference filter, Ge: germanium collimating lens. Half-wave plates are not shown. Inset: The spectrum of the ps-NIR pulses measured using an FTIR spectrometer. The FWHM is 0.50 nm.

The portion of the laser beam reflected off the ultra-narrow bandpass filter was split into two parts (cf. BS in Fig. 1) to pump and seed an optical parametric pre-amplifier (i.e., OPA1). 7.1-µJ, 514-nm pump pulses were produced for OPA1 via second-harmonic generation (SHG) in an anti-reflection (AR)-coated 2-mm-long β-BaB$_2$O$_4$ (BBO) crystal (type I, θ = 23.4°) starting from 18-µJ, 1-µm pulses. The remaining 1-µJ portion of the reflected beam was used for generating > 1100-nm supercontinuum seed pulses in a 6-mm-long, uncoated YAG crystal (SCG). The 514-nm pump and the seed pulses were sent to a beam combining dichroic mirror (DM1) and focused into a 2-mm-long, AR-coated BBO crystal in a collinear geometry, i.e., cut for type I phase matching at θ = 24°. The spectral range between 1100 and 1300 nm with an incident $1/e^2$ beam diameter of 410 µm was amplified in OPA1 as a signal wave to an average power of 38-58 mW, depending on the center wavelength.

The so-produced tunable signal pulses seeded the second OPA stage (OPA2), which was pumped by the remaining 34-µJ portion of the 1028-nm pump beam. OPA2 was based on an uncoated, 5-mm-long LGS crystal cut for type I phase matching in the x-z plane at θ = 48.2°. The pump and seed ($1/e^2$ diameter of 890 µm) beams were collinearly combined at a dichroic mirror (DM2), and the seed pulses were amplified at a peak incident pump intensity of 57 GW/cm$^2$. The focus of both the pump and seed pulses was placed in front of the LGS crystal, providing a divergent beam incident on the crystal to mitigate self-focusing [12]. The generated MIR idler pulses passed through two custom-made dichroic mirrors (DM3), afterward were collimated with an $f$ = 200 mm AR-coated germanium lens (Ge), and were compressed by transmission through an AR-coated germanium slab.

## 3. Characterization of laser pulses

### 3.1 Narrowband ps-NIR

Out of the 0.86 µJ/pulse transmitted through the ultra-narrow bandpass filter, an energy of 0.82 µJ was available for experiments. The spectrum was measured using a commercial Fourier-transform optical spectrum analyzer and is shown in the inset in Fig. 1. The FWHM was measured to be 0.50 nm, which corresponds to a transform-limited duration of 3.1 ps for Gaussian laser pulses, resulting in a spectral resolution of 4.7 cm$^{-1}$ in our

VSFG measurements. An overall throughput of 4.3% was achieved in the combined filtering and beam transport process.

3.2 Recycled pump

1028-nm fundamental pulses with 19 µJ/pulse reflected off the ultra-narrow bandpass filter were available for pumping and seeding OPA1. The pulses directly before and after spectral filtering were characterized using a home-built SHG frequency-resolved optical gating (SHG-FROG) device equipped with a 10-µm thin, type I BBO and group-delay-optimized 50% beam splitters specified for the 600-1500 nm range. Despite the fact that group delay dispersion (GDD) was not a design consideration for the bandpass filter coating, the component showed exceptionally high performance also in reflection. Figures 2 and 3 show the measured and retrieved SHG-FROG traces and the reconstructed temporal and spectral intensity profiles of the fundamental 1028-nm beam before and after the spectral filtering, respectively. The retrieved pulse duration of the fundamental beam was 186 fs, which is in good agreement with the factory test specification of the laser. The pulse duration obtained for the back-reflected beam was identical to the unfiltered duration within the error budget of the FROG retrieval. The reconstructed spectral phase showed strong modulation only in the immediate vicinity of the filtering wavelength, and the resulting temporal pedestal, extending beyond 1 ps, was weak. The π phase jump at the filter wavelength may be an artifact due to the ambiguity of SHG-FROG at multiples of π [13].

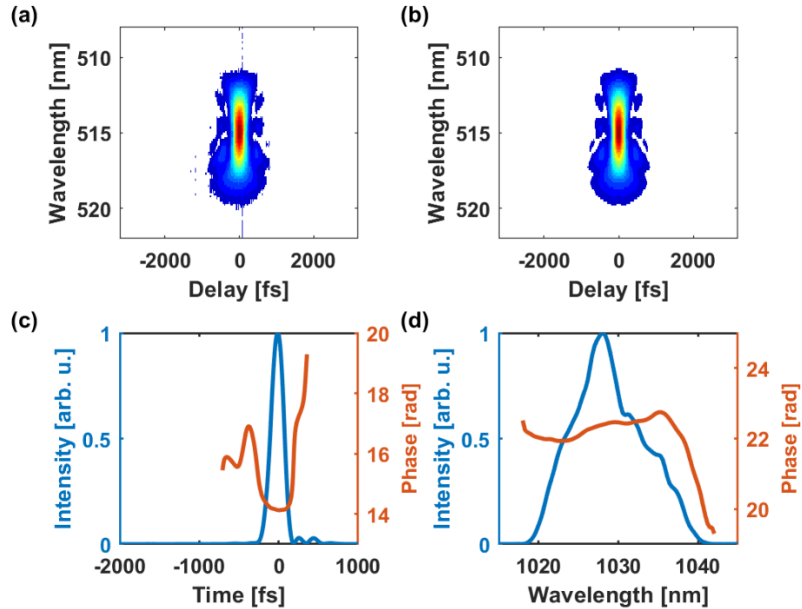

**Fig. 2.** SHG-FROG data obtained for the 20-µJ arm of the 1028-nm pulses before spectral filtering. The FROG-error of the reconstruction for a grid size of 256 × 256 points was 0.0015, and the retrieved pulse duration was 186 fs.

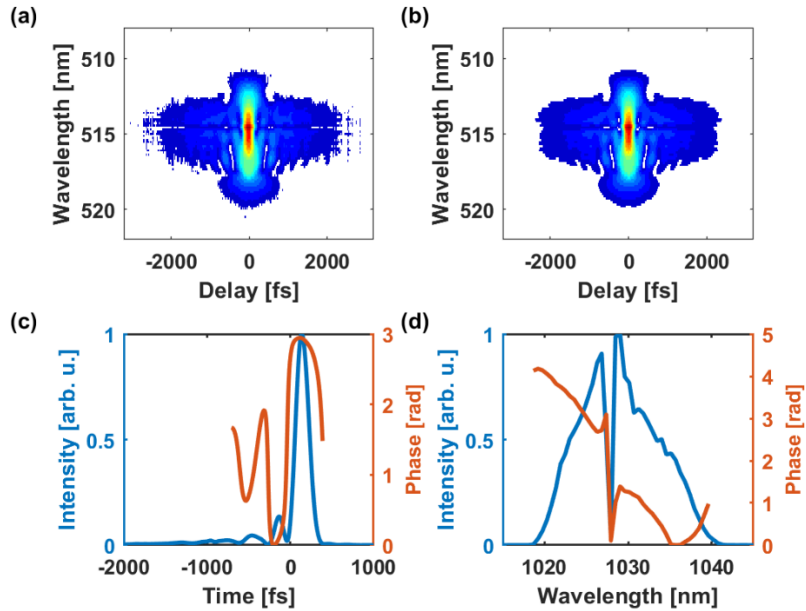

**Fig. 3.** SHG-FROG data obtained for the recycled 1028-nm pulses, i.e., reflected off the filter. The FROG-error of the reconstruction for a grid size of 256 × 256 points was 0.0044, and the retrieved pulse duration was 182 fs.

3.3 Output of OPA1: tunable near-infrared pulses

The signal beam of the BBO OPA front-end falling in the range of 1100-1300 nm served as the seed of the LGS OPA (OPA2). SHG-FROG characterization was done at 1230, 1180, and 1160 nm, where OPA1 delivered average powers of 48, 54, and 58 mW, respectively. As a representative example, Fig. 4 shows the measured and retrieved SHG-FROG traces and the reconstructed temporal and spectral intensity profiles obtained at a center wavelength of 1180 nm, which served as the seed in OPA2 for the generation of ca. 8.0-µm (1250-cm$^{-1}$) idler pulses. Qualitatively similar FROG-reconstructed data were obtained at the other two wavelengths. While the second-harmonic pump pulses were derived from a pulse with a spectral hole, the supercontinuum pulses in the 1100-1300 nm range and the corresponding signal output pulses do exhibit a smooth, continuous spectrum. Fitting a fourth or fifth-order polynomial on the retrieved spectral phase, the absolute values of the extracted GDD and third-order dispersion (TOD) amount to $(1.47\pm0.07)\times10^3$ fs$^2$ and $(1.35\pm0.15)\times10^4$ fs$^3$, respectively. The reconstructed pulse duration was 158 fs, while the transform-limited pulse duration amounted to 20.0 fs.

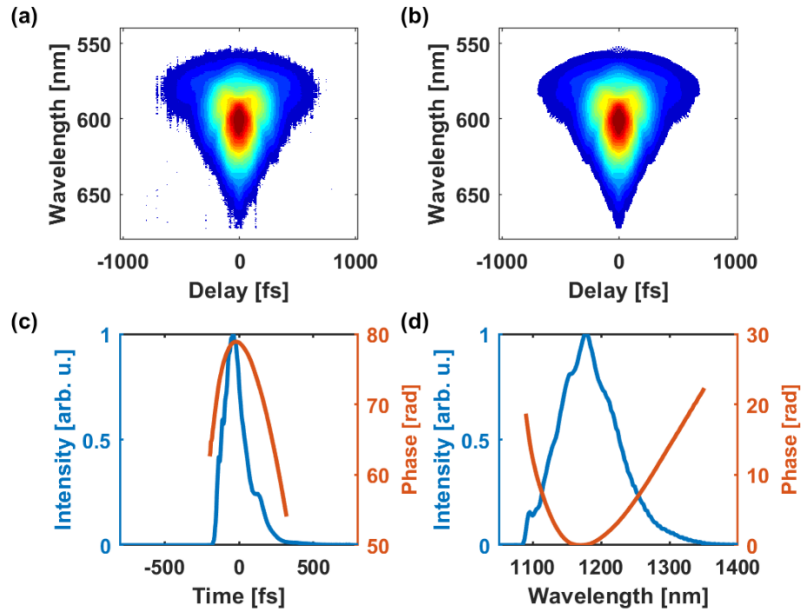

**Fig. 4.** SHG-FROG data obtained for 1180-nm signal pulses from OPA1. The FROG-error of the reconstruction for a grid size of 256 × 256 points was 0.0037, and the retrieved pulse duration was 158 fs.

Figure 5 shows the results of SHG-FROG characterization for the idler output pulses from OPA1 centered at 900 nm, corresponding to the 1180-nm signal pulses shown in Fig. 4. Fitting a fourth or fifth-order polynomial on the retrieved spectral phase, the absolute values of the extracted GDD and TOD amount to $(3.69\pm0.01)\times10^2$ fs$^2$ and $(6.7\pm0.1)\times10^3$ fs$^3$, respectively. The reconstructed pulse duration was 28.2 fs, while the transform-limited pulse duration amounted to 21.5 fs. The idler pulses were much closer to the transform limit since they were generated with negative GDD and were partially compressed when propagating through transmissive N-BK7 and fused silica optics.

The spectral phase data obtained through SHG-FROG retrieval suggest that both the signal and idler pulses feature smooth, trivial polynomial spectral phases, which can be compensated in a straightforward way if necessary. When seeding OPA2 using the signal pulses, the generated MIR idler pulses are expected to show similarly high spectral quality.

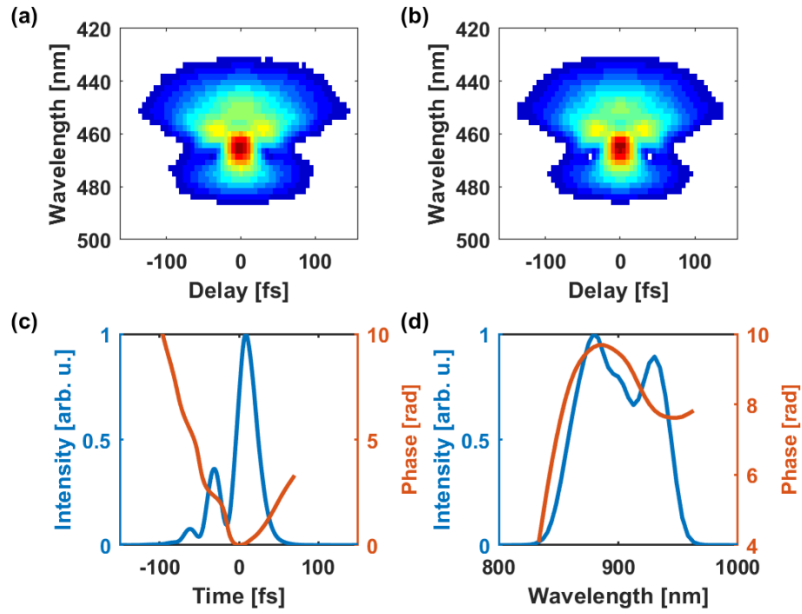

**Fig. 5.** SHG-FROG data obtained for 900-nm idler pulses from OPA1, corresponding to the 1180-nm signal pulses in Fig. 4. The FROG-error of the reconstruction for a grid size of 64 × 64 points was 0.0051, and the retrieved pulse duration was 28 fs.

3.4 Output of OPA2: tunable mid-infrared pulses

After collimation with the Ge lens, the tunable MIR idler pulses were sent to the input of a VSFG spectrometer [14], where the available average powers were measured (Fig. 6, top panel). Under optimal broadband phase matching conditions of LGS, i.e., signal and idler group velocity matching, the broadest spectral bandwidth of FWHM = 1.6 µm was obtained at 7.8 µm, corresponding to 260 cm$^{-1}$ and a transform-limited duration of 57 fs (cf. Fig. 6, top panel, grey circled symbol). However, the maximum power of 91 mW, i.e., 0.91 µJ, was reached at 7.4 µm (Fig. 6, top panel, green circled symbol). The corresponding practical overall pump-to-idler energy conversion efficiency was 2%, defined as the ratio between the idler pulse energy available for VSFG measurements and the pump power incident at the LGS input surface. The LGS OPA and the VSFG spectrometer were purged by filtered dry air to avoid absorption lines due to atmospheric $CO_2$ and $H_2O$ in the MIR spectrum.

The bottom graph of Figure 6 displays the MIR idler spectra obtained by sum-frequency mixing the MIR pulses with the narrowband 1028-nm pulses (ps-NIR) at a planar gold interface using our home-built VSFG spectrometer [14] which was adapted to accommodate the longer VSFG wavelengths. Wavelength tuning was realized by changing the LGS crystal tilt angle and optimizing the delay between the seed and pump pulses. The longest center wavelength reached 9.5 µm, which was limited by the infrared absorption cut-off of LGS [15]. On the other hand, the shortest available center wavelength was 4.8 µm, limited by the short wavelength cut-off of the Ge longpass filter that was used to separate the MIR pulses from the NIR pump and signal pulses.

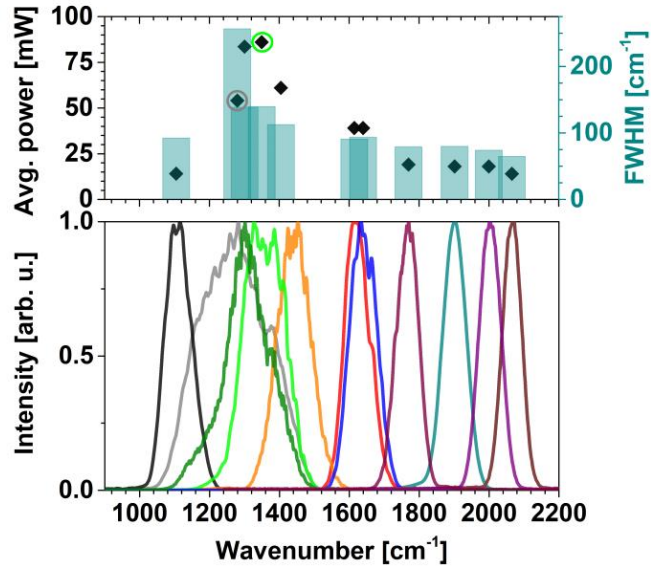

**Fig. 6.** Tunability of the LGS OPA. Top: average power of the idler pulses (black symbols, left scale) as a function of center wavelength across the tuning range and the corresponding spectral FWHM of the MIR idler pulses (green bars, right scale). Bottom: representative MIR idler spectra across the tuning range.

Figure 7(a) shows the energy conversion efficiency as a function of pump power obtained at a center wavelength of 7.4 µm, i.e., 1350 cm$^{-1}$. The corresponding MIR spectrum and average power are marked by light green in Fig. 6. Saturation was reached at a pump power of 3.38 W.

Since the spectral FWHM bandwidth of the front-end, 700 cm$^{-1}$, is well beyond that of the maximum bandwidth of the MIR output from a 5-mm-long LGS crystal, we also tested a 2-mm LGS sample. The spectra obtained with the 2- and 5-mm long crystals at a center wavelength of 7.4 µm (i.e., the highest power delivered by the 5-mm sample) are shown in Fig. 7(b). The longer crystal generated a FWHM bandwidth of 139 cm$^{-1}$ (cf. light green line in Fig. 6(a) and 7(b)), while the thinner crystal delivered 255 cm$^{-1}$ (blue line in Fig. 7(b)). While the use of the thinner crystal resulted in a significantly broader bandwidth, the output power was only 17% of the power delivered by the 5-mm sample.

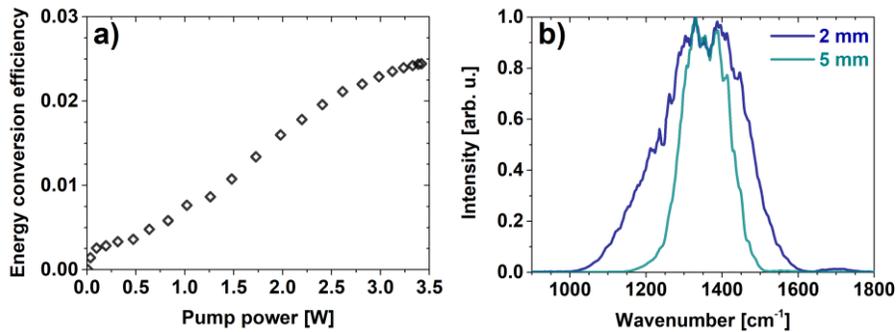

**Fig. 7.** (a) Practical pump-to-idler energy conversion efficiency obtained at 1350 cm$^{-1}$, i.e., 7.4 µm, as a function of 1-µm pump power. (b) MIR spectrum obtained with 2-mm and 5-mm long LGS crystals.

## 4. Proof-of-concept: observing a phospholipid monolayer at the air-water interface by VSFG spectroscopy

To benchmark the performance of our VSFG spectrometer utilizing the two new beamlines, i.e., the 1028-nm ps-NIR beam obtained through bandpass filtering and the broadband MIR beam obtained through the use of the novel front-end, we measured the VSFG spectrum of a phospholipid monolayer at the air-water interface. The results were compared with those obtained in similar measurements performed in the past using our standard setup based on a narrowband VIS beam at 514 nm and a broadband MIR beam delivered by an OPA that was seeded directly with supercontinuum pulses without any bandpass filtering involved [14]. In our VSFG spectrometer, the MIR and VIS/NIR pulses were focused on the interface with angles of incidence of 58° and 68° relative to the surface normal, respectively. In the case of our standard setup, the pulse energies at the sample were 0.25 µJ and 4.15 µJ for the MIR and the 514-nm beams, respectively. In comparison, with our new setup, 0.3 µJ and 0.81 µJ of pulse energy was used in the MIR and at 1028 nm, respectively. The VSFG spectra were collected in an SSP polarization combination (i.e., the letters denote the polarization of the VSFG, VIS/NIR, and MIR beams, respectively) with a 320-mm imaging spectrometer equipped with a Peltier-cooled CCD. Each raw VSFG spectrum was background corrected, then frequency calibrated by using a 50-µm polystyrene film as the calibration standard and normalized with a reference MIR spectrum obtained on a bare gold surface [16].

A Langmuir monolayer of 1,2-dipalmitoyl-sn-glycero-3-phosphocholine (DPPC) on water subphase was prepared by dropping 5 µl stock solution of 1mg/ml DPPC (grade >99%, Avanti Polar Lipids, Inc.) dissolved in 9:1 chloroform:methanol. This method resulted in a monolayer on top of ultrapure water (18.2 MΩ cm of resistivity) with a surface pressure of ~35 mN/m, serving as a simple membrane model since this value is close to the surface pressure measured in cellular membranes [17].

To demonstrate the capability of the new spectrometer, an average of ten VSFG spectra of the DPPC monolayer at the air-water interface were collected with both VSFG spectrometers in the spectral range of 1000-1400 cm$^{-1}$ (i.e., 7.1-10 µm). The results are shown in Fig. 8. At a 5.1 times lower average power of the new upconversion beam and a 25% shorter acquisition time, the spectrum acquired using the new setup is almost identical to that obtained by our standard setup. Thanks to the unusually narrow bandwidth of the bandpass filter, the spectral shoulder in Fig. 8 at ca. 1060 cm$^{-1}$ is well resolved. Some loss in signal-to-noise ratio (S/N) is observed with the new setup compared to our old setup, likely due to the much lower power of the bandpass-filtered 1028-nm beam compared to the 514-nm beam obtained through high-efficiency nonlinear spectral compression. Nevertheless, regarding the quality of the acquired VSFG spectrum of DPPC at the air-water interface, the combination of our resolution, S/N, and MIR spectral bandwidth still compares favorably to that of standard broadband VSFG spectroscopy [18, 19]. More importantly, using the long upconversion wavelength at 1028 nm helps overcome detrimental electronic resonance and re-absorption effects in certain sensitive biomolecules by moving the SFG wavelengths to the NIR.

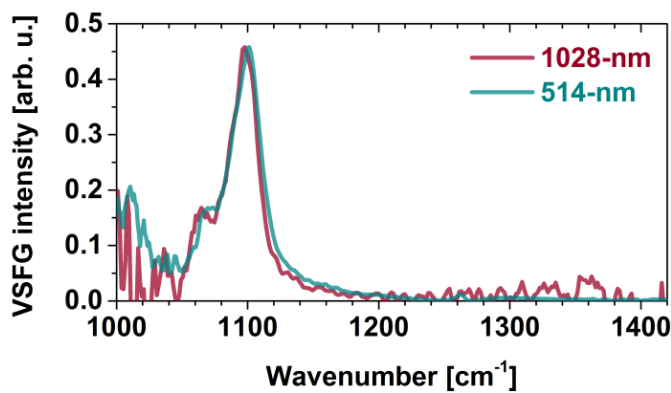

**Fig. 8.** Average of 10 normalized VSFG spectra of a DPPC monolayer at the air-water interface acquired in SSP polarization combination with 1028-nm ps-NIR pulses in 30 s (red) and 514-nm VIS pulses in 40 s (green).

## 5. Conclusion

Various types of nonlinear vibrational spectroscopy (e.g., certain implementations of SRS, CARS, VSFG) derive their high spectral resolution from narrow bandpass-filtering of femtosecond NIR laser pulses [20, 21]. Since the beam rejected by the filter is commonly discarded, the corresponding procedure is inherently very lossy. We demonstrated that contrary to the prevailing belief, 180-fs, 1.03-µm laser pulses reflected off an ultranarrow bandpass filter can be utilized for driving an ultrafast NIR OPA, providing both the seed pulses through supercontinuum generation and the pump pulses. We showed that the OPA constructed in this manner can serve as a pre-amplifier, boosting the seed pulse energy of a MIR OPA operating in the molecular fingerprint region. We also demonstrated that the MIR beam, together with the narrowband, filtered 1028-nm beam, can efficiently drive a VSFG spectrometer and high signal-to-noise ratios are achievable at short acquisition times and at a spectral resolution ($< 5$ cm$^{-1}$) that is beyond that obtained in typical bandpass filtering schemes ($\geq 10$ cm$^{-1}$). Since the sum-frequency spectrum now falls into the NIR, the scheme enables the acquisition of vibrational spectra without detrimental electronic resonance and re-absorption of the upconverted sum-frequency beam triggering photo-physical processes in sensitive biological samples. Shifting the VSFG signal well above 700 nm can also be utilized to access molecular interfaces buried under biological layers. Employing the presented VSFG spectrometer layout can open up new perspectives in surface/interface sciences, especially for bioapplications.

**Funding.** Deutsche Forschungsgemeinschaft (DFG) (GSC 1013 SALSA and PE 607/14-1); European Union's Horizon 2020 research and innovation programme under Grant Agreement No. 871124 Laserlab-Europe.

**Acknowledgments.** Z.H. acknowledges funding by a Julia Lermontova Fellowship from DFG, No. GSC 1013 SALSA. We thank Alluxa, Inc., for sharing data and details about the design considerations of the ultra-narrow bandpass filter used in this work.

**Disclosures.** The authors declare no conflicts of interest.

**Data availability.** Data underlying the results presented in this paper are available upon reasonable request.